\documentclass[showpacs,aps,superscriptaddress]{revtex4}
\usepackage{amsmath}
\usepackage{times}
\usepackage{amssymb}
\usepackage{graphics}
\usepackage{float}
\usepackage{color}
\usepackage[utf8]{inputenc}
\usepackage[english]{babel}
\def\<{\langle}
\def\>{\rangle}
\begin{document}
\today
\title{
Anomalous Fourier's law and long range correlations in a 1D non-momentum
conserving mechanical model}
\author{A. Gerschenfeld}
\author{B. Derrida}
\affiliation{ Laboratoire de Physique Statistique, Ecole Normale
Sup\'erieure, UPMC Paris 6,Universit\'e Paris Diderot, CNRS,
24 rue Lhomond, 75231 Paris Cedex 05 - France}
\author{J. L. Lebowitz} 
\affiliation{Departments of Mathematics and Physics, Rutgers University, 110 Frelinghuysen Road, Piscataway, NJ 08854, USA}

\keywords{non-equilibrium systems, large deviations, current
fluctuations}
\pacs{02.50.-r, 05.40.-a, 05.70 Ln, 82.20-w}

\begin{abstract}
We study by means of numerical simulations the velocity reversal model, a
one-dimensional mechanical model of heat transport introduced in 1985 by
Ianiro and Lebowitz. Our numerical results indicate that this model, although
it does not conserve momentum, exhibits an anomalous Fourier's law similar to
the one s previously observed in momentum-conserving models. This
is contrary to what is obtained from the solution of the Boltzmann equation
(BE) for this system. The pair correlation velocity field also looks very
different from the correlations usually seen in diffusive systems, and shares
some similarity with those of momentum-conserving heat transport models.\\
\ \\
keywords: current, anomalous Fourier law, hard particle gas
\end{abstract}

\keywords{non-equilibrium systems,  current
fluctuations,  anomalous Fourier's law}

\maketitle

\date{\today}

\section*{Introduction}

Understanding the steady state of a system in contact with two heat baths at
unequal temperatures is a central question to the theory of
non-equilibrium systems\cite{BLR}. With very few exceptions (such as zero
range processes\cite{Evans2000,Evans2005}), non equilibrium systems exhibit
long range correlations in their steady
state\cite{spohn1983,Schmitz1985,DKS,OJS}. These correlations have been
studied theoretically by various approaches and measured experimentally.

For a one dimensional diffusive system of size $L$, for instance a stochastic
lattice gas, these long range correlations take the following scaling
form\cite{spohn1983,BDGJL09,DLS,D07}:
\begin{equation}
\langle A(r_1) A(r_2) ... A(r_n)
\rangle_c = {1 \over L^{n-1}} F_n \left( {r_1 \over L}, ... {r_n \over L}
\right)\,,\label{diff_corr}
\end{equation}
(where $A(r)$ is an observable at position $r$ such as
the density or the energy) when the distances between the positions $r_i$ are
macroscopic (i.e. $ L \sim |r_i - r_j| \gg 1$).

The macroscopic fluctuation theory developed by Bertini et
al.\cite{BDGJL1,BDGJL2} allows one to write down the general equations
satisfied by the scaling functions $F_n$\cite{BDGJL09}. These partial
differential equations are usually difficult to solve. The expressions of the
$F_n$ are however known in a number of examples\cite{spohn1983,BDLW,BD10}, where they have been obtained
either from exact solutions of microscopic models or by integrating the
partial differential equations derived from the macroscopic fluctuation
theory. Diffusive systems are also known to satisfy Fourier's law,
meaning that, for large system sizes, the steady state flux $ \langle J
\rangle$ of energy through a system of size $L$ scales like $L^{-1}$:
\begin{equation} \langle J_L \rangle = {1 \over L}  G(T_a,T_b)\,,
\label{Fourier}
\end{equation}
where $G(T_a,T_b)$ is a function of the temperatures $T_a$ and $T_b$ of the
heat baths at the two extremities of the system, which maintain it out of
equilibrium; the function $G(T_a,T_b) $ vanishes linearly with the difference
$T_a-T_b$ so that the flux becomes a gradient.

Over the last 15 years\cite{LLP,LLP2,Dhar,LW,BBO,BBO2,MDN,DeN,ILOS} it has
been realized on the basis of numerical simulations that
one-dimensional mechanical systems which conserve momentum do not
satisfy Fourier's law. Instead, they exhibit a power law decay, called
anomalous Fourier's law, of the average flux $J$ with system size:
\begin{equation}
 \langle J_L \rangle = L^{\alpha -1}  G(T_a,T_b)\,.
\label{anomalous}
\end{equation}
So far, the exponent $\alpha$ has not been determined analytically for any
microscopic model with non-quadratic interactions; however,
numerical\cite{GNY,DN} and analytic\cite{CP,LLP3,Dhar2,BBO} calculations,
based on mode coupling theory or on other approaches\cite{NR,LuS}, indicate
that, depending on the type of the non-linearities and on the accuracy of the
simulations, $\alpha$ can take values ranging between $0.2$ and $0.5$. The
main two systems for which this anomalous heat conduction has been observed
are:
\begin{itemize}
  \item the Fermi-Pasta-Ulam model, a chain of $N$ harmonic oscillators with
  an additional cubic (FPU-$\alpha$) or quartic (FPU-$\beta$) interaction
  potential\cite{LLP,DLRP,LuS} . For the latter case, most estimates indicate that
  $\alpha\sim0.25$;
  \item a one-dimensional gas of $N$ hard point particles with elastic
  collisions\cite{DN} (when the particles are identical, the collisions simply exchange
  the velocities of the incoming particles: hence the system has the same
  transport properties as an ideal gas, i.e. a ballistic transport with
  $\alpha=1$). For the collisions to be non-trivial, one choice is to consider
  particles of alternating masses 1 and $m_2\neq1$ (2-mass model). In this
  case, it is commonly found that $\alpha\sim0.33$\cite{GNY}.
  
\end{itemize}
In comparison with diffusive systems, for which (\ref{Fourier}) holds, the
main feature of these models is that their dynamics conserve momentum, and it
is believed that this property plays a key role in the divergence of the heat
conductivity\cite{PC}. Note, however, that there is no proof for this
behavior: the argument of \cite{PC} is not correct, although its conclusion 
is consistent with observations.

In this article, we consider a simpler model originally introduced in
\cite{lebowitz}, the velocity reversal model, in order to better understand
which features could be at the origin of the anomaly. This model is, like the
1D hard-particle gas, a system of $N$ free-moving particles undergoing
collisions. The collisions are, however, not elastic, but instead are given by
the following simple rule:
\begin{itemize}
  \item when two particles collide with velocities of opposite signs, $v_{i+1} 
  < 0 < v_i$, their two velocities get reversed ( $v_i \to -v_i$, 
  $v_{i+1} \to -v_{i+1}$ at the collision);
  
  \item when they collide with velocities of the same sign, the particles
  simply pass each other ( $v_i \to v_{i+1}$ and $v_{i+1}\to v_i$ if the
  particles are kept ordered from left to right).
\end{itemize}

In contrast with the Fermi-Pasta-Ulam chain and the hard-particle gas, these
dynamics do not conserve momentum. It should be noted, however, that the
absolute values of the velocities themselves are conserved, leading in general
to more conserved quantities than in standard momentum-conserving models. One
should also note that, when the velocities are reversed at independent
exponential times instead of at collisions, the system becomes
diffusive\cite{lebowitz}.

Here, the $N$ particles of the velocity reversal model are in a
one-dimensional box of length $L=N$ between two boundaries which play the role
of heat baths. Whenever a particle hits a boundary, its velocity is changed as
if the particle was reinjected instantaneously, with its velocity thermalized
by the reservoir. The reservoirs are thus described by the velocity p.d.f. of
these "reinjected particles", $\rho_a(v)$ and $\rho_b(v)$. An obvious choice
is to consider Maxwellian reservoirs at temperatures $T_a$ and $T_b$, with
velocity densities
\begin{equation}\label{gaussien}
   \rho_a(v) = \theta(v) {v\over T_a}e^{-{v^2\over 2 T_a}} \mbox{ and }
  \rho_b(v) = \theta(-v) {|v|\over T_b}e^{-{v^2\over 2 T_b}}\mbox{ , with }
  \theta(v) = \left\lbrace \begin{array}{l}
    1 \mbox{ for } v > 0\\
    0 \mbox{ for } v \leq 0
  \end{array}\right.\,.
\end{equation}
We will study below this "Maxwellian case" for $T_a = 4$ and $T_b = 1$. For a
system which thermalizes well, one would expect other choices of these
reservoir velocity densities to only affect small regions near the boundaries,
but not to influence the macroscopic behavior of the system. For the velocity
reversal model, however, the conservation of the absolute values of the
particle velocities prevents thermalization from taking place: for instance,
one can easily see that, for another choice of reservoirs, where
particles are always reinjected with velocities $v_a$ from the left reservoir
and $v_b$ for the right reservoir,
\begin{equation}\label{2vitesses}
\rho_a(v) = \delta(v-v_a) \mbox{ and } \rho_b(v) = \delta(v+v_b)\,,
\end{equation}
the only possible velocities inside the system are $\pm v_a$ and $\pm v_b$. We
will study this "two-speed case" with $v_a = 2$ and $v_b=1$.

\bigskip

In a first part, we show, by studying the steady-state current of systems with
$125\leq N \leq 8000$ particles, that, both in the Maxwellian and the
two-speed case, the velocity-reversal model exhibits anomalous Fourier's law,
in contrast to the prediction of the associated Boltzmann
equation\cite{lebowitz}.

We then present, in a second part, measurements of the steady-state
correlation functions of the momentum density for Maxwellian and two-speed
systems of $100\leq N\leq 400$ particles, as well as for a 2-mass
hard-particle gas with $m_1=1$ and $m_2=1.6$ for comparison: they indicate a size
dependence very different from (\ref{diff_corr}).

\section{Anomalous Fourier's law for the velocity reversal model} 
\label{sec:anomalous_fourier_s_law_for_the_velocity_reversal_model}

The main advantage of the velocity reversal model described above, compared to
the hard-particle gas for instance, is that its associated Boltzmann equation
can be solved analytically, thanks to the simpler form of its collision term.
By solving this Boltzmann equation in the steady state with the appropriate
boundary conditions, the average current between the reservoirs can be
predicted, as done in \cite{lebowitz}. The simplest case is the two-speed
system (\ref{2vitesses}) with velocities $v_a$ and $v_b$, for which the
Boltzmann equation relates the four densities $f_{\pm a}(x,t)$ and $f_{\pm b}
(x,t)$ of the particles with respective velocities $\pm v_a$ and $\pm v_b$,
with $0\leq x\leq L=N$ the space coordinate :

$$\left\lbrace \begin{array}{l}
(\partial_t + v_a \partial_x)f_{+a}= (v_a+v_b)(f_{+b}f_{-a}-f_{-b}f_{+a}) \equiv 
A(x,t)\,;\\
(\partial_t - v_a \partial_x)f_{-a}= -A(x,t)\,;\\
(\partial_t + v_b \partial_x)f_{+b}=-A(x,t) \,;\\
(\partial_t - v_b \partial_x)f_{-b}= A(x,t) \,.\\
\end{array}\right. $$
\begin{figure}[h]
  \includegraphics{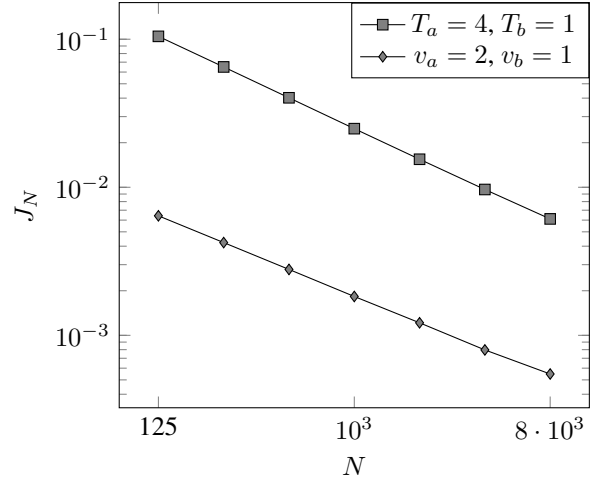}
  \caption{Steady-state energy current $J_N$ for the velocity-reversal model
  with $125\leq N\leq 8000$ particles, in the Maxwellian case with $T_a=4$ and
  $T_b = 1$ (above) and in the two-speed case with $v_a = 2$ and $v_b=1$
  (below). In both cases, the current decreases as a non-integer power of 
  the system size : over the range of sizes we considered, our data are
  well-fitted by $J_N \propto N^{-0.69}$ in the Maxwellian case
  and $J_N \propto N^{-0.60}$ in the two-speed case.
  }
  \label{joel_cur}
\end{figure}
\begin{figure}[h]
  \includegraphics{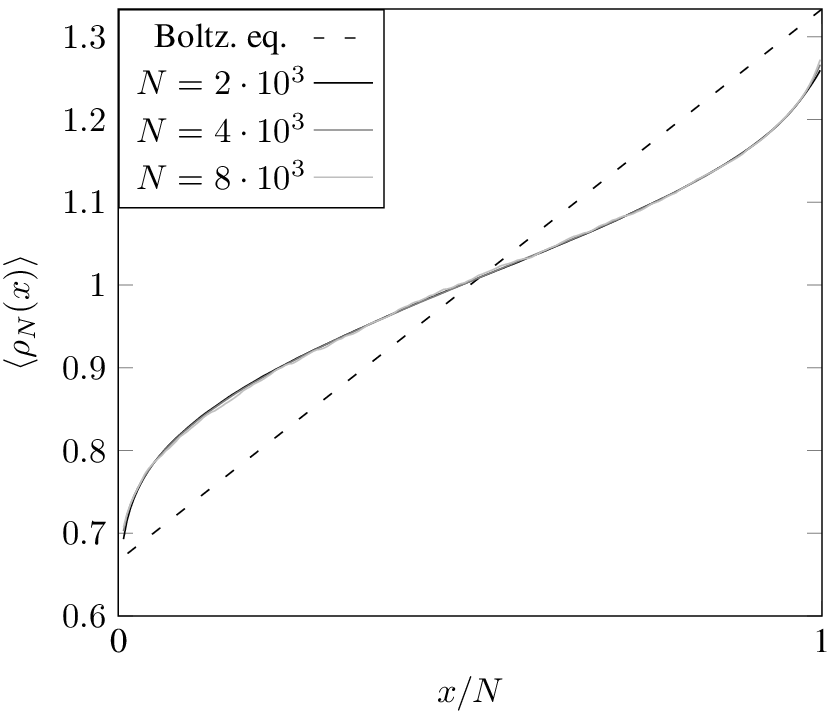}
  \includegraphics{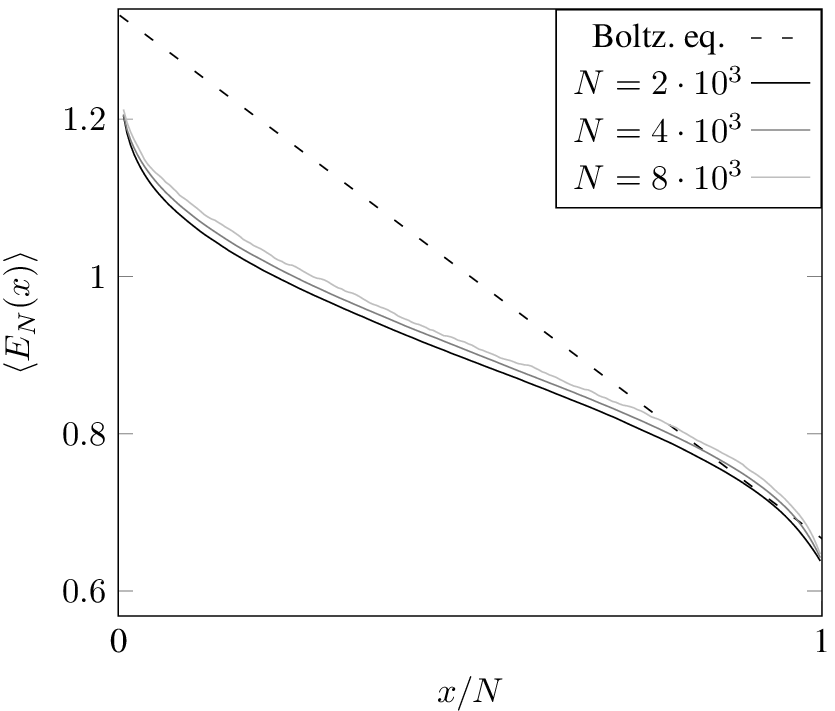}  
  \caption{Average density (left) and energy (right) profiles for the
  two-speed velocity reversal model between reservoirs for $v_a=2$ and $v_b=1$
  and for system sizes $N=2\cdot 10^3$, $4\cdot 10^3$ and $8\cdot 10^3$
  (continuous lines). According to the Boltzmann equation, these profiles
  (given by $f_{+a}+f_{-a}+ f_{+b}+f_{-b}$ and ${v_a^2\over2}(f_{+a}+f_{-a}) +
  {v_b^2\over2}(f_{+b}+f_{-b})$ respectively) should be linear in $x$.}
  \label{2v_prof}
\end{figure}\\
Setting the time derivatives to $0$, these equations can be
solved\cite{lebowitz} to determine the steady state of the two-speed system:
$$\left\lbrace \begin{array}{l}
f_{+a}(x) = {v_b\over v_a + v_b}\left(1-x +{x\over N+1}\right)\\
f_{-a}(x) = {v_b\over v_a + v_b}{N\over N+1}(1-x)\\
f_{+b}(x) = {v_a\over v_a + v_b}{N\over N+1}x\\
f_{-b}(x) = {v_a\over v_a + v_b}\left(x +{1-x\over N+1}\right)\,,
\end{array}\right.$$
yielding linear profiles for the four densities as well as a steady-state
current $J = {v_a v_b(v_a-v_b)\over 2(N+1)}$ satisfying Fourier's
law(\ref{Fourier}). By a similar calculation, the Boltzmann equation also
predicts that the energy current should also satisfy Fourier's law in the
Maxwellian case.

We studied numerically the Maxwellian (\ref{gaussien}) and the two-speed
(\ref{2vitesses}) velocity reversal models for $125\leq N \leq 8000$
particles, with $T_a = 4$, $T_b = 1$ in the first case and $v_a = 2$, $v_b=1$
in the second. Because the true steady state is unknown, our measurements were
performed by starting the system in an
equilibrium configuration at $T=2$ in the Maxwellian case, and by choosing the
particle velocities uniformly at random among $\pm v_a$, $\pm v_b$ in the
two-speed case.

We then let the system evolve from that initial state, and sampled the time
evolution of $J_N$, $\rho_N(x)$ and $E_N(x)$ : $J_N$ was measured by a time
average of the instantaneous energy flux, $J_N = {1\over2N}\sum_{i=1}^N
v_i^3$, while $\rho_N(x)$ and $E_N(x)$ were measured by counting the particles
and energy within the boxes $k\leq x < k+1$ for $k = 0,..,N-1$. We then took
the long-time limits of these measured quantities to estimate their values in
the steady state.

As shown in figure \ref{joel_cur}, our data indicate that $J_N$ decreases like
a power law of $N$ for both choices of reservoirs for the system sizes we
considered. They are consistent with (\ref{anomalous}), with $\alpha \sim 0.3$
for the Maxwellian reservoirs and $\alpha \sim 0.4$ for the two-speed
reservoirs. In the latter case, the density and energy profiles $\rho_N(x)$
and $E_N(x)$ shown in figure \ref{2v_prof} also differ noticeably from the
linear profiles predicted by the Boltzmann approach, at least for the sizes we
were able to study.
\section{Long-range correlations in anomalous systems} 
\label{sec:long_range_correlations_in_anomalous_systems}

The main assumption in the derivation of the Boltzmann equation is that
there are no correlations between the velocities of particles entering a
collision in which their velocities are reversed. As it fails to predict the
anomalous Fourier's law of figure \ref{joel_cur}, it is interesting to
investigate the form of the steady-state correlations for the velocity
reversal model.

\begin{figure}[h]
  \includegraphics{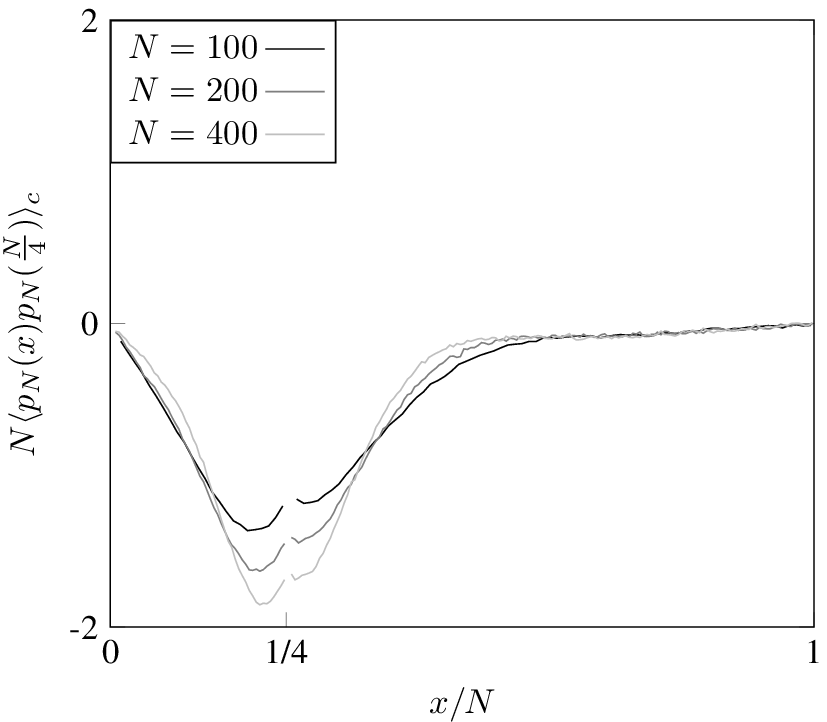}
  \includegraphics{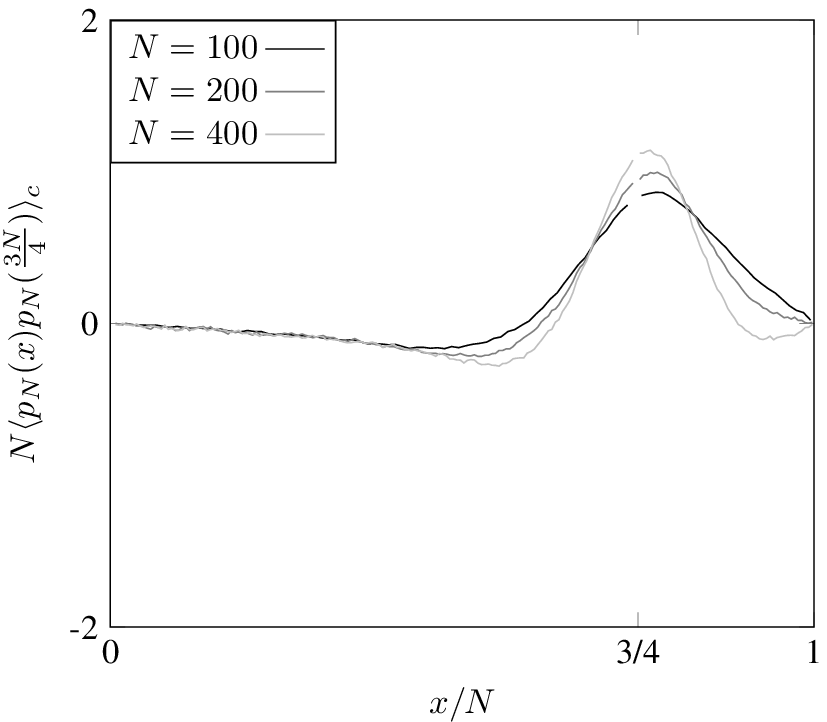}
  \caption{Two-point correlation function $\langle p(x)p(y)\rangle_c$ of the
  momentum field $p_N(x)$ for the velocity reversal model in the Maxwellian
  case (\ref{gaussien}) with $T_a=4$ and $T_b=1$, for $y={N\over4}$ (left) and
  $y={3N\over4}$ (right).}
  \label{g_cv}
\end{figure}

\begin{figure}[h]
  \includegraphics{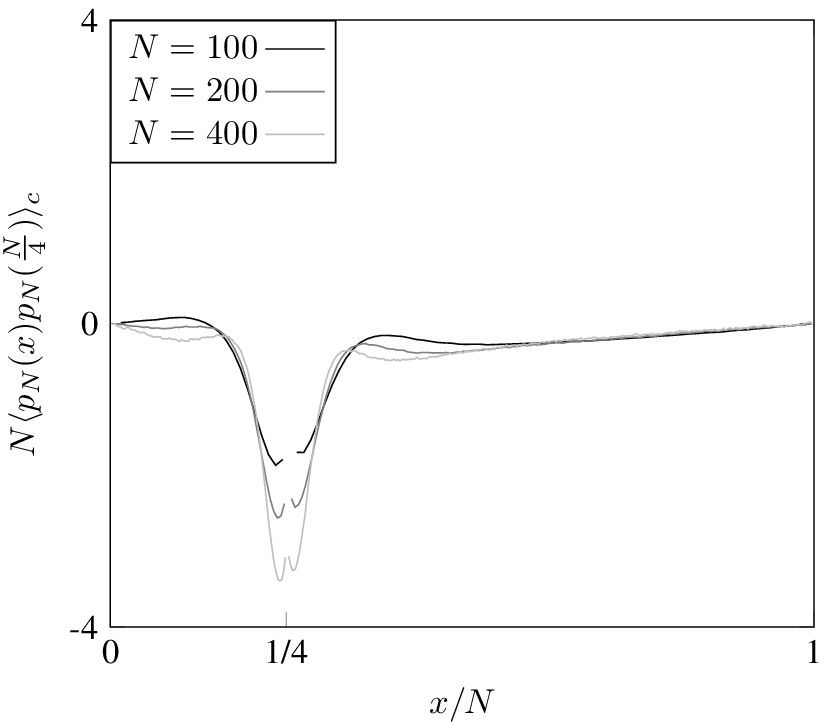}
  \includegraphics{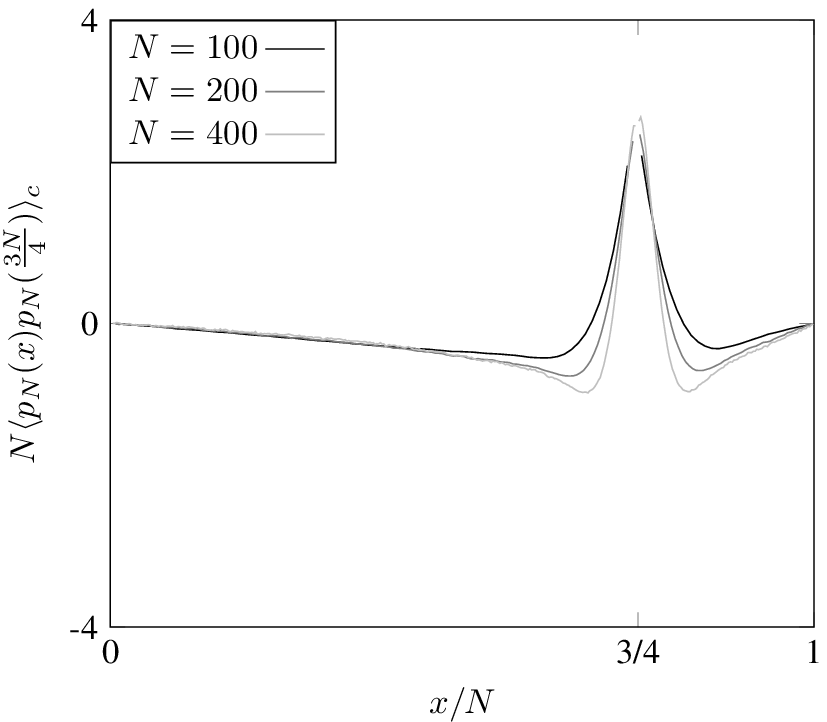}
  \caption{Two-point correlation function $\langle p(x)p(y)\rangle_c$ of the
  momentum field $p_N(x)$ for the velocity reversal model in the two-speed
  case (\ref{2vitesses}) with $v_a=2$ and $v_b=1$, for $y={N\over4}$ (left)
  and $y={3N\over4}$ (right).}
  \label{2v_cv}
\end{figure}

We measured numerically the steady-state two-point correlation functions of
the momentum field, $\langle p(x)p(y)\rangle_c$, for the velocity reversal
model in the Maxwellian (fig. \ref{g_cv}) and in the two-speed (fig.
\ref{2v_cv}) cases, for systems of $100$ to $400$ particles.

As in the previous section, we approximated the momentum density function
$p_N(x)$ by its discretization over the boxes $k \leq x < k+1$ for $0\leq k <
N$. We took the stationary-state correlations $\langle p(x)p(y)\rangle_c$,
which we measured for $y={N\over4}$ and $y={3N\over4}$, to be
the long-time limits of their time evolutions for a system started in the
arbitrary initial state described in the previous section.

For diffusive systems, one would expect $\langle p(x)p(y)\rangle_c$ to follow
a scaling of the form (\ref{diff_corr}): $\langle p(x)p(y)\rangle_c = {1\over
N} F_2({x\over N}, {y\over N})$, except for $x=y$ with a variance $\langle
p(x)^2\rangle_c$ of order 1. For the velocity reversal models of figures
\ref{g_cv} and \ref{2v_cv}, we observed $\langle p(x)^2\rangle_c$ to be of
order $1$: however, when measuring $\langle p(x)p(y)\rangle_c$ for $y=N/4$
and $y=3N/4$, as in figures \ref{g_cv} and \ref{2v_cv}), we observed an
additional anomalous gowth in $\langle p(x)p(y)\rangle_c$ for $x$ close to
$y$ in a region growing with system size, but becoming narrower on a
macroscopic scale.

The sign of $\langle p(x)p(y)\rangle_c$ seems to be always negative for $y$
closer to the left (warmer) reservoir and positive for $y$ closer to the
right, colder reservoir. As shown in figures \ref{g_cv} and \ref{2v_cv}, where
our data are multiplied by the system size $N$, the correlations seem to
decay slower than $1/N$ as $N$ grows, over a region which seems to grow slower
than the system size.

\begin{figure}[h]
  \includegraphics{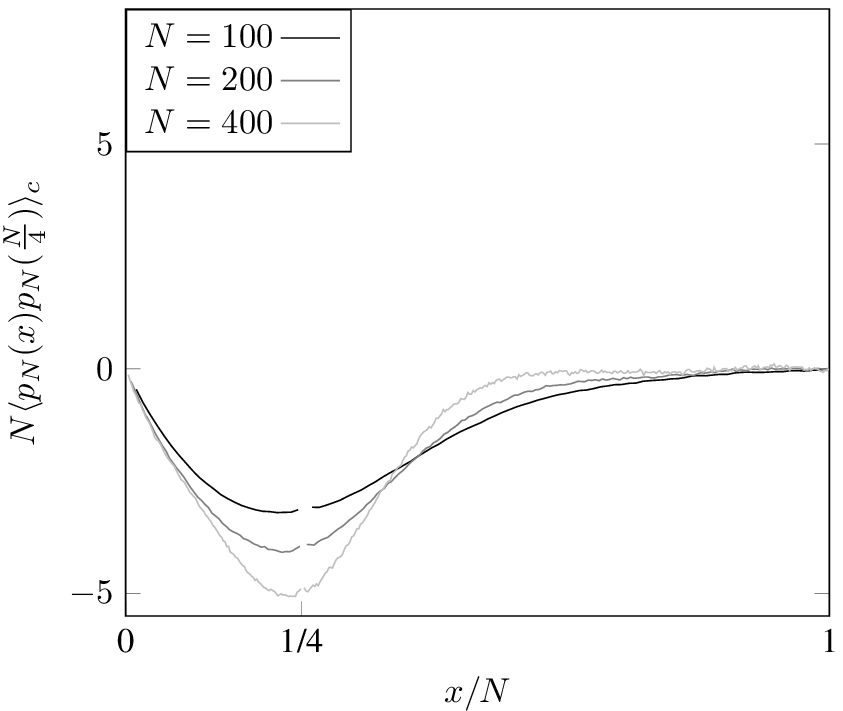}
  \includegraphics{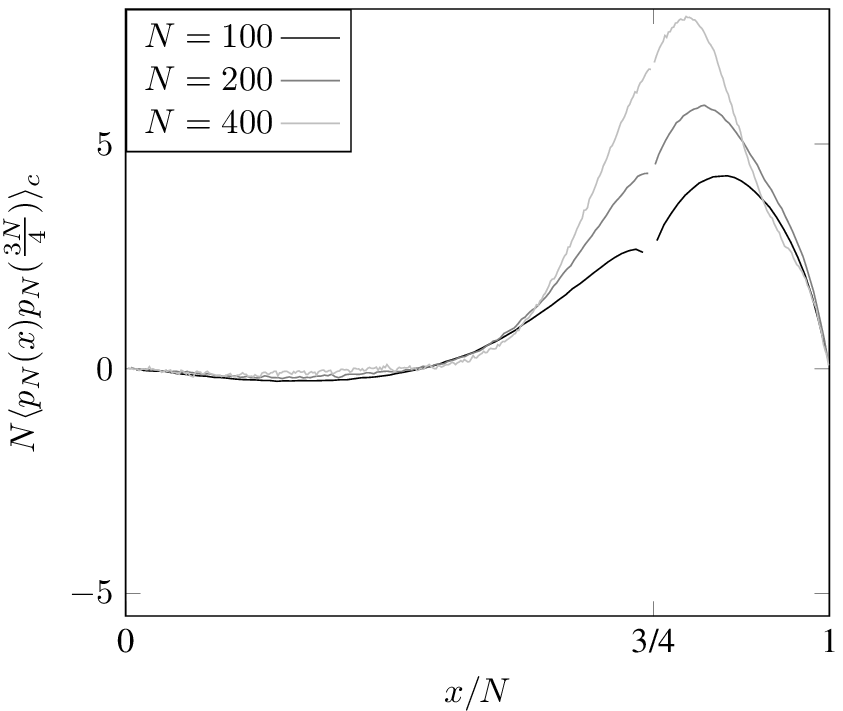}  
  \caption{Two-point function $\langle p(x)p(y)\rangle_c$ of the momentum
  field $p_N(x)$ for the 2-mass hard-particle gas with $m_2 = 1.6$ between
  reservoirs at $T_a=4$ and $T_b=1$, for $y={N\over4}$ (left) and
  $y={3N\over4}$ (right).}
  \label{2m_cv}
\end{figure}

In order to compare the behavior of these models with those of usual
momentum-conserving models, we also measured the two-point correlation
function of the momentum for a hard-particle gas of particles of alternating
masses $1$ and $m_2 = 1.6$ between Maxwellian reservoirs at $T_a = 4$ and $T_b
= 1$ : we observed a similar growth of the correlation
functions at intermediate scales (fig. \ref{2m_cv}). 

\begin{figure}[h]
  \includegraphics{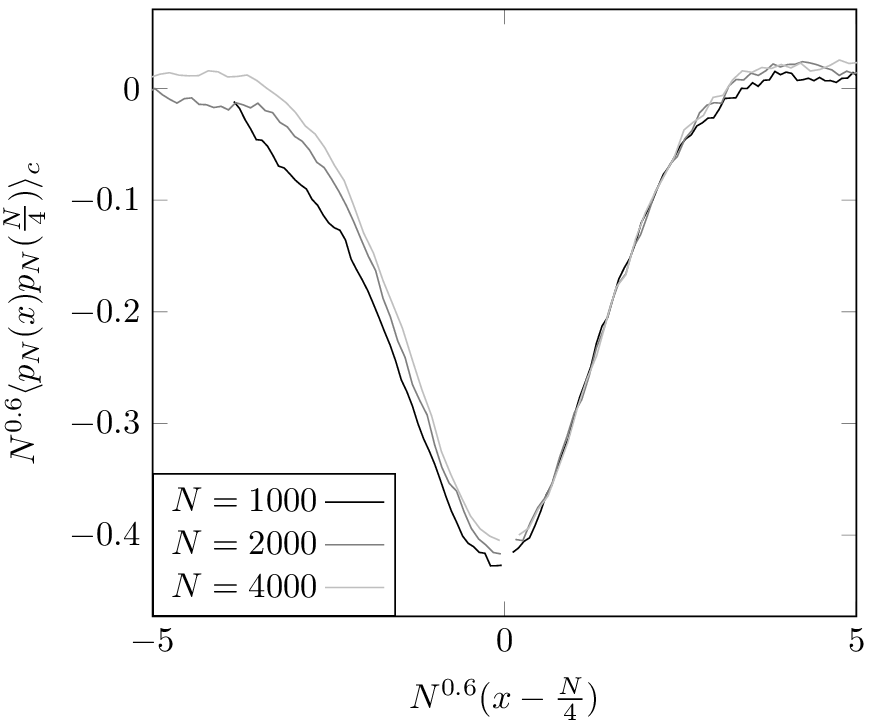}
  \includegraphics{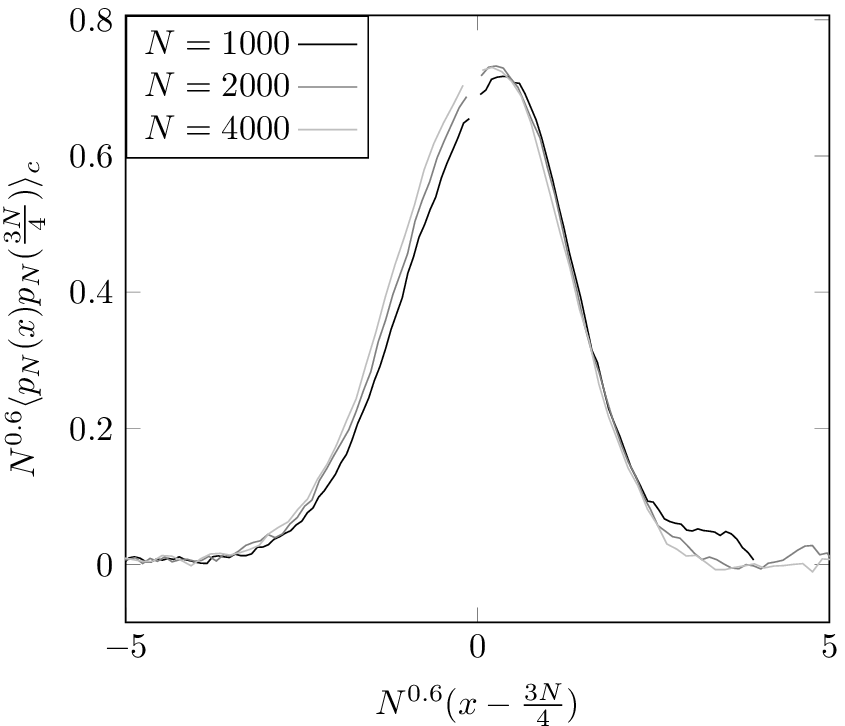}
  \caption{Rescaling of the two-point function of the momentum field $\langle
  p_N(x) p_N(y)\rangle_c$, for $y={N\over4}$ (left) and $y={3N\over4}$
  (right), for the 2-mass hard particle gas with $m_2=1.6$ between reservoirs
  at temperatures $T_a=4$ and $T_b=1$. }
  \label{scaling}
\end{figure}

For this hard-particle gas (fig. \ref{2m_cv}), due to a faster relaxation to
the steady state, we were able to simulate much larger system sizes than for
the velocity reversal model: at these sizes, $10^3 \leq N \leq 4\cdot 10^3$,
the anomalous growth of $ \langle p_N(x)p_N(y)\rangle_c$ of figure \ref{2v_cv}
is compatible with a scaling of the form
$$ \langle p_N(x)p_N(y)\rangle_c \propto {1\over N^{0.6}} F_2\left( x
- y\over N^{0.6}\right)\,,$$
as shown in figure \ref{scaling}. The sizes we could reach for the
velocity-reversal model of figures \ref{g_cv} and \ref{2v_cv} did not exhibit
such a clear scaling form.

For the velocity-reversal model as well as for the 2-mass hard particle gas,
we also measured the correlations of the energy field $E_N(x)$, which
exhibited a behavior very similar to the one shown above for the momentum
field.

\section*{Conclusion} 
\label{sec:conclusion}

The numerical simulations presented in this work show that the
velocity-reversal model exhibits an anomalous Fourier's law, although it
does not conserve momentum.
Its density and energy profiles (fig. \ref{2v_prof} ) differ
noticeably from those predicted by the Boltzmann equations. For the sizes we
could achieve, however, they have not yet converged: this might question
whether our data showing the anomalous Fourier's law (fig. \ref{joel_cur})
correspond to the asymptotic regime.

Within the system sizes we could reach, the two-point function of the
momentum also looks very different from those of diffusive systems: they seem
to be concentrated on a mesoscopic scale, much larger than the microscopic
scale, but smaller than the system size. Within this mesoscopic region, the
decay of the correlations seems to be anomalous as well, i.e. seems to be a
non-integer power of the system size.

We have checked that the pair correlations seem to have a rather similar
behavior for the 2-mass hard particle gas. This leads us to believe that the
shape of the correlation functions we have observed is probably another
signature of the anomalous Fourier's law: it would therefore be interesting to
see whether these anomalous correlations are also present for other models
known to exhibit an anomalous heat conductivity, such as the Fermi-Pasta-Ulam
chain.

Looking at higher-order correlations would require major numerical efforts.
They would however be very interesting to measure in order to guess what would
replace (\ref{diff_corr}) for systems exhibiting anomalous Fourier's law.
 
The present work is purely numerical in nature. It would of course be very
interesting to see what the existing theories explaining the anomalous
Fourier's law would predict for the pair correlation functions we have
measured. For diffusive systems, the scaling (\ref{diff_corr}) of the
correlations with system size is closely linked to the scaling form of the
large deviation function of the system's density profile\cite{D07,BDGJL09}:
what such a scaling form would look like for systems exhibiting anomalous
Fourier's law is also an interesting open issue. 

\acknowledgements{We would like to thank J. Lukkarinen, H. Spohn, and H. van
Beijeren for useful discussions. J.L. thanks the IHES and the IHP for
their hospitality during this work. The work of J.L. was supported by NSF grant
DMR-044-2066 and by AFOSR grant AF-FA9550-10.}

\end{document}